\documentclass[pre, twocolumn, superscriptaddress, showpacs]{revtex4}

\usepackage{graphicx}
\usepackage{amsmath}
\usepackage{amssymb}
\usepackage{mathrsfs}
\usepackage{epstopdf}

\begin{document}

\title{Size effects in dislocation depinning models for plastic yield}

\author{Zoe Budrikis}
\affiliation{ISI Foundation, Via Alassio 11/c, 10126 Torino, Italy}
\author{Stefano Zapperi}
\affiliation{CNR-IENI, Via R. Cozzi 53, 20125 Milano, Italy}
\affiliation{ISI Foundation, Via Alassio 11/c, 10126 Torino, Italy}

\begin{abstract}
Typically, the plastic yield stress of a sample is determined from a stress-strain curve by defining a yield strain and reading off the stress required to attain it. However, it is not \textit{a priori} clear that yield strengths of microscale samples measured this way should display the correct finite size scaling. Here we study plastic yield as a depinning transition of a $1+1$ dimensional interface, and consider how finite size effects depend on the choice of yield strain, as well as the presence of hardening and the strength of elastic coupling. Our results indicate that in sufficiently large systems, the choice of yield strain is unimportant, but in smaller systems one must take care to avoid spurious effects.
\end{abstract}

\pacs{62.20.F-, 64.60.an}

\maketitle

\section{Introduction}
A complete understanding of microscale plasticity is important both for fundamental and technological reasons, and much experimental and theoretical effort has been expended on this problem in recent years. A key result that has emerged --- and that is not intuitively obvious from our everyday experience of classical bulk plasticity --- is that on the microscale plastic flow is subject to size effects, with recent experiments showing that in submicron samples, the yield stress increases with decreasing sample size~\cite{Uchic2004}. However, the origins of these finite size effects are not clear, and many mechanisms have been proposed (for reviews, see Refs~\cite{Zaiser2006, Greer2011, Kraft2010}).

In typical experimental studies, the yield stress is determined from measured stress-strain curves by defining a yield strain $\gamma_Y$, typically a few percent, and reading off the stress required to attain such a strain. This procedure suffers from the fact that measured stress-strain curves are not smooth and the transition to plastic yield is generally preceded by power law distributed avalanches~\cite{Dimiduk2006}. This raises the question of how the yield strain should be selected to ensure that the measured yield stress is not systematically over- or under-estimated in a way that affects finite size scaling.  Furthermore, it is not \textit{a priori} clear that one can compare experimental results in which different threshold strains have been used, making it difficult to synthesize results from different studies.

In this paper, we approach this problem from a theoretical perspective, by studying numerical simulations of $1+1$ dimensional depinning at zero temperature. This is the simplest model for plastic	 yielding in a crystal: the interface represents a dislocation, driven through a random potential generated by disorder and other dislocations in the sample. Depinning has previously been studied in a variety of contexts, using both numerical and analytical techniques~\cite{Feigelman1983, Bruinsma1984, Nattermann1992, Narayan1993, Leschhorn1993, Amaral1995, Rosso:2001, Vannimenus2001, LeDoussal2002, Rosso2003, Bolech:2004, Fedorenko2006, Middleton2007}, and it is well-known that there exists a critical driving force $f_c$, above which the interface is depinned and has finite velocity always, and below which the interface will always come to a stop at long times. This transition is analogous to plastic yielding, in which stresses above a finite threshold are able to induce large strain increases that ideally are limited only by sample failure.

Like other phase transitions, the depinning transition is subject to finite size effects, so that the depinning force in a system of linear size $L$ has the asymptotic form
\begin{equation}
\label{FiniteSize}
f_c(L) = f_{\infty} + a L^{-1/\nu}
\end{equation}
where $f_{\infty}$ is the depinning force in the limit $L\to\infty$ and $\nu$ is the finite size exponent, determined by the universality class of the model. A key question that this work aims to address is whether the observed finite size scaling is affected by how the depinning transition is measured.

\begin{figure}
\begin{center}
\includegraphics[width=0.9\columnwidth]{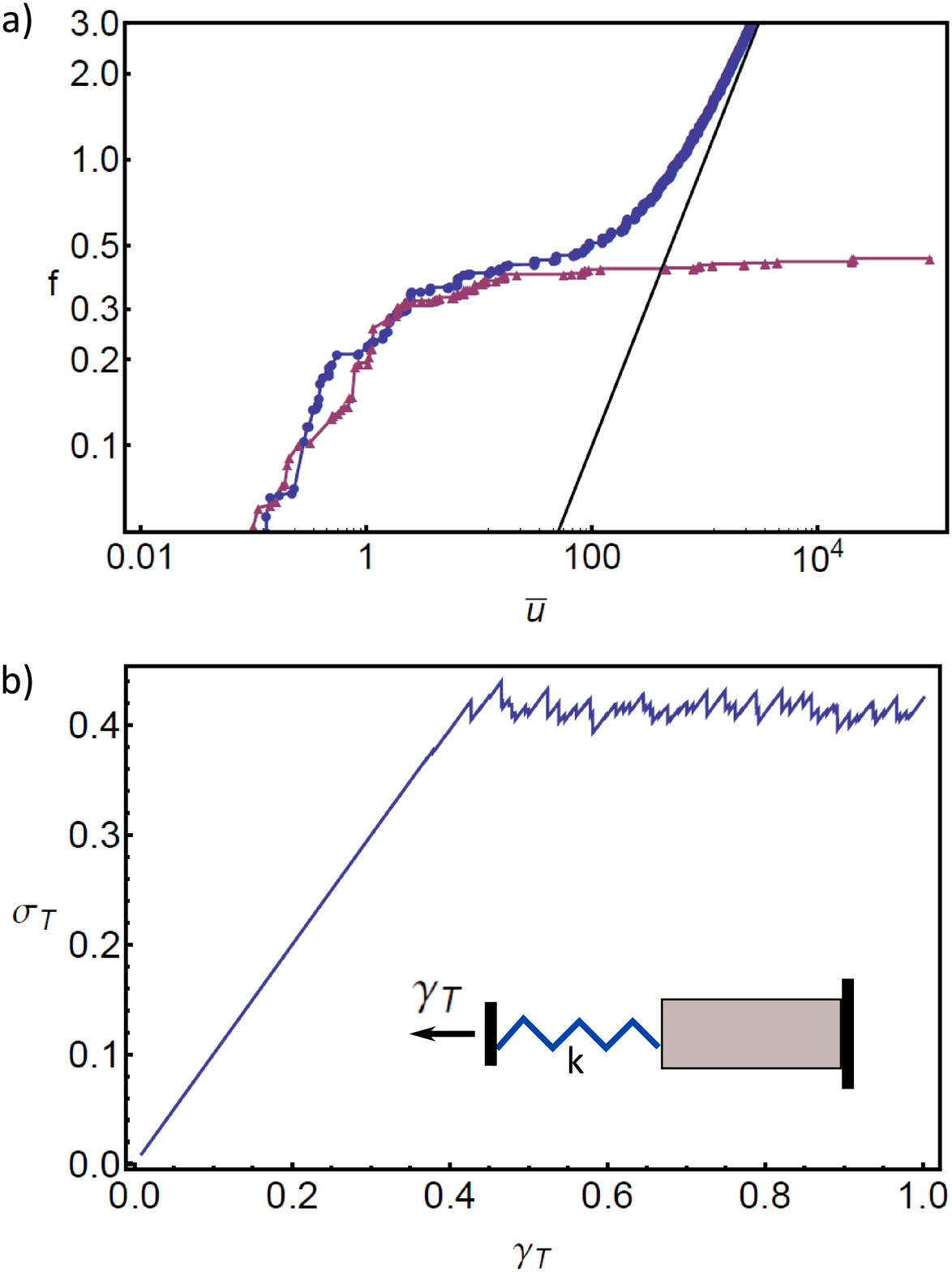}
\end{center}
\caption{(a)Typical ``stress-strain'' (driving force, $f$, \textit{vs} mean interface displacement, $\bar{u}$) curves from simulations of the quenched Edwards-Wilkinson equation, with (circles) and without (triangles) hardening. The black solid line indicates the $f\propto \bar{u}$ behavior of the model with hardening as $\bar{u}\to\infty$ (see Section~\ref{hardening}). Depinning forces are determined by selecting a target value of $\bar{u}$ and reading off the driving force required to attain this target. (b) Data from a hardening simulation recast as a strain controlled experiment with machine stiffness $k$, using the transformations $\sigma_T = f-k\bar{u}/L^2$, $\gamma_T = f/k$, as described in Section~\ref{hardening}. The yield stress, $\sigma_{T, y}$ is determined by selecting a yield strain $\gamma_{T,y}$ and reading off the stress required to attain it.}
\label{StressStrain}
\end{figure}

In particular, we can imitate the experimental procedure for measuring yield stress by mapping our simulations of interface depinning onto plots of stress \textit{vs} plastic strain from stress controlled experiments. The motion of a dislocation with Burgers vector $b$ over distance $x$ in a system of size $L^2$ results in a strain increase of $\gamma=b x/L^2$, and the driving force is proportional to an external stress. We can therefore consider a plot of the mean position of the interface \textit{vs} driving force to represent plastic strain \textit{vs} stress, as illustrated in Fig.~\ref{StressStrain}(a). In analogy to the approach taken in experimental analysis, we define a target displacement $\Delta=\gamma L^2/b$ to define the onset of depinning, and study the effects of varying this target. In what follows, we will measure lengths in units of $b$.

We note that this approach to measure the depinning threshold raises another issue. It has previously been pointed out~\cite{Bolech:2004, Duemmer2005, Fedorenko2006} that in a simulation of a system of finite size $L \times M$ with periodic boundary conditions in both directions, a ``natural'' choice of $M$ is $M \sim L^{\zeta}$, where $\zeta$ is the interface roughness exponent and $L^{\zeta}$ gives the interface width at criticality. In this way, one ensures that for all $L$, the system contains the same number of critical configurations. On the other hand, our target scales as $L^2$, so although it is natural for the plasticity scenario at hand, the system aspect ratios increase with $L$. We will see that this affects the distribution of measured depinning forces, as anticipated~\cite{Bolech:2004, Fedorenko2006}.

The depinning models can also be related to strain-controlled experiments,  by driving the interface via a spring of stiffness $k$, and adjusting the externally-applied stress so that the total strain --- the elastic strain of the spring, partially relieved by the plastic deformation arising from interface motion --- increases smoothly. The set-up and a typical stress-strain curve are illustrated in Fig.~\ref{StressStrain}(b).  Again in this context, we study the effect of the choice of yield strain on measured yield stress, and discuss how these effects can be related to strain hardening in a stress controlled experiment.

The outline of the paper is as follows. In Section~\ref{EW}, we use the quenched Edwards-Wilkinson model for interface depinning as a case study for the role of target choice on measured properties, considering not only the finite size scaling of the mean depinning force but also how the choice of target affects its whole distribution. 
The ideas developed in Section~\ref{EW} are independent of universality class, which we demonstrate in Section~\ref{MFSect} via results from the Kardar-Parisi-Zhang (KPZ) model and a mean field model. Finally, in Section~\ref{hardening} we discuss two closely-related extensions to the quenched Edwards-Wilkinson model, namely strain hardening and the above-mentioned strain controlled experiments. We show that finite size effects in these models are more complex than in the other models we study, depending on both the finite size scaling of the depinning force and also the hardening amplitude or stiffness of the coupled spring.

\section{Effects of target choice: the quenched Edwards-Wilkinson equation}
\label{EW}
We start by considering a simple linear model for the interplay between elasticity, random pinning and an external drive, namely, the quenched Edwards-Wilkinson model~\cite{Feigelman1983, Bruinsma1984}. This will allow us to discern the effect of the choice of target interface displacement in a straight-forward manner, giving results that will aid the interpretation of more complex models that follow.

In general, an overdamped elastic interface characterized by height function $u(x)$ and subject to a random potential and a uniform external drive $f$ is described by overdamped equations of motion
\begin{subequations}
\begin{align}
\frac{1}{\Gamma}\dot{u}(x) &= f_{\rm tot}(x)\\
 &=f + f_{\rm el}(x, u(x)) - \eta(x, u(x)),
\end{align}
\end{subequations}
where the damping constant $\Gamma$ is hereafter set to unity.
The self-interaction of the interface, $f_{\rm el}$, promotes it to be flat, and may be local or long-range, depending on the model at hand. The self-interaction is in competition with the quenched disorder $\eta$, which causes the interface to prefer to pass through sites with low potential energy. In the quenched Edwards-Wilkinson model that we study in this Section, the elastic force at position $x$ on the line is given by $\nabla^2 u(x)$.

In this work, we study a cellular automaton version of this model, by discretizing $x$ and $u$. The elastic force is given by $u_{x+1}+u_{x-1}-2u_x$. We allow only forward motion of the interface: at each simulation step, a site $x$ may increase $u_x$ by one unit if the total force acting on it is positive, or remain stationary if the force is zero or negative. The random potential is implemented by drawing a random number for each $x, u_x$. In general, these are from a uniform distribution over $[0, 1]$, but in Section~\ref{MFSect} we will also study Gaussian distributed disorder, because when all sites are coupled to one another, the disorder distribution can be important~\cite{Duxbury2001}. We use periodic boundary conditions in the $x$ direction, and open boundary conditions in the $u$ direction. 

In order to generate ``stress-strain'' curves in a stress-controlled simulation, we start with a flat interface with $u_x=0$ everywhere and $f=0$. We increase the drive $f$ adiabatically, by finding the site $x^{*}$ with the smallest required driving force for motion, and increasing $f$ so that $x^{*}$ can move. This may trigger an avalanche via changes in $f_{\rm el}$, and we update sites in parallel until no more can move. We record the value of $f$ and the mean position of the interface attained after it has stopped moving, and then increase the drive again. 
As mentioned in the Introduction, the record of driving forces and mean interface positions is the ``stress-strain'' curve, since the motion of a dislocation by distance $x$ increases the plastic strain by $x/L^2$ (we work in units where the Burgers vector $b$ is unity). We continue to increase the drive until the mean interface position passes the target $\gamma L^2$. The driving force required for the target to be attained is denoted $f_c$, and for simplicity we will call it a ``depinning'' threshold, even if this is not strictly accurate.

\begin{figure}
\begin{center}
\includegraphics[width=0.9\columnwidth]{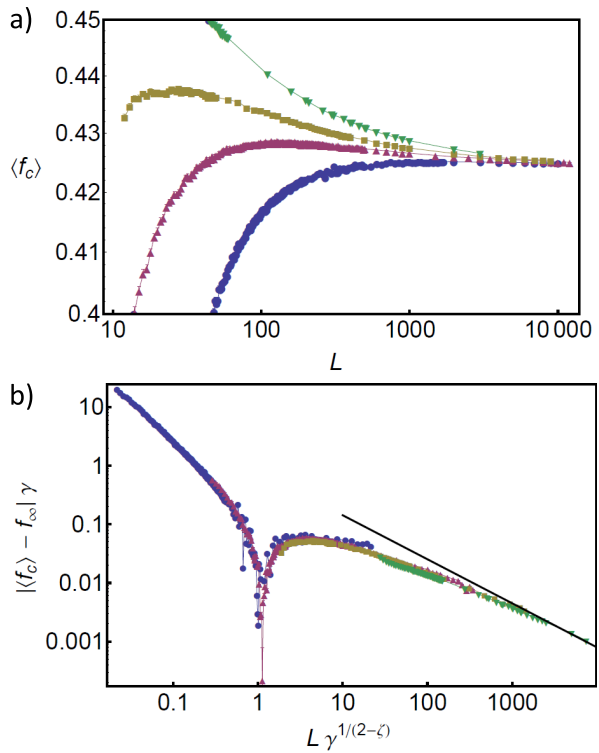}
\end{center}
\caption{(a) Mean measured depinning force $\langle f_c \rangle$ plotted against system linear size $L$, from simulations of the quenched Edwards-Wilkinson equation. The target strains used are $\gamma=0.01$ (circles), $0.07$ (triangles), $0.25$ (squares) and $2$ (inverted triangles). For all targets, $f_c$ approaches the same value as $L\to\infty$, but the behavior for smaller $L$ depends on $\gamma$.  Error bars (typically smaller than the symbols) represent the standard error. (b) The same data, collapsed using Eq.~\eqref{FiniteSize} and \eqref{TargetGreaterWidth}. The black solid line is proportional to $L^{-0.75}$, which is the expected finite size scaling for the model for large $L$.}
\label{qEW}
\end{figure}

\begin{figure}
\begin{center}
\includegraphics[width=\columnwidth]{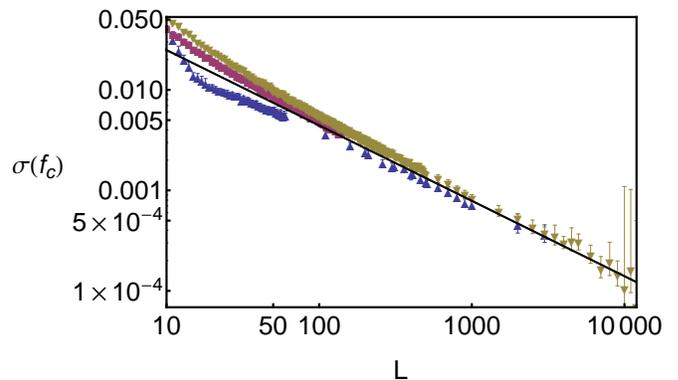}
\end{center}
\caption{Standard deviation in the measured depinning force, $\sigma(f_c)$, plotted against system linear size $L$. For sufficiently large targets, $\sigma(f_c) \sim L^{-0.75}$, as indicated by the solid black line. The target strains used are $\gamma=0.07$ (circles), $0.15$ (triangles), $2$ (squares). Because the distributions of $f_c$ are not Gaussian (see Fig.~\ref{fcDistribution}), the error bars have been estimated using the method in Ref.~\cite{Bonett2006}.}
\label{qEWStdDev}
\end{figure}

Figure~\ref{qEW}(a) shows the mean force required for the interface to attain the target mean position $\gamma L^2$, as a function of system linear size $L$, for a range of $\gamma$ values between $0.01$ and $2$. In the large-$L$ limit, we find that $f_c(L)$ approaches $f_{\infty}=0.424$ according to $L^{-1/\nu}$, with $1/\nu=0.75$, in agreement with previous studies of finite size scaling~\cite{Bolech:2004, LeDoussal2002}. However, for small $L$, a distinct $\gamma$ dependence becomes apparent: for small $\gamma$, $f_c$ initially grows with $L$, but for large $\gamma$, $f_c(L)$ decreases monotonically.

This result can be understood by considering that there are two length scales in the direction of interface propagation. The first is the interface width at depinning, which is given by $w L^{\zeta}$, with the roughness exponent $\zeta=1.25$ for the quenched Edwards-Wilkinson equation~\cite{Leschhorn1993, Rosso2003}  and $w$ a constant. The second length scale is our choice of target, which is $\gamma L^2$. The two length scales are equal when
\begin{equation}
\label{TargetGreaterWidth}
L^{2-\zeta}=\frac{w}{\gamma},
\end{equation}
giving a crossover length scale $L_c=(w/\gamma)^{1/(2-\zeta)}$.

If $L<L_c$, the target is shorter than the interface width at criticality, and the simulation is halted during the early stages of interface motion, before the depinning transition can occur. In this regime, increasing $L$ --- and thereby the target --- should increase the final driving force applied before the simulation is stopped. On the other hand, for $L>L_c$, we can expect the interface to undergo the depinning transition, and the standard finite size scaling of Eq.~\eqref{FiniteSize} should hold. At $L_c$, there is a transition from the former behavior to the latter. However, if $\gamma$ is sufficiently large, the target is always larger than the interface width, that is, $L_c \to 0$ as $\gamma \to \infty$.

To verify this interpretation, we collapse the data from simulations with different $\gamma$ values. From Eq.~\eqref{TargetGreaterWidth}, we see that we must rescale $L \to L \gamma^{-1/(2-\zeta)}$. Then from the finite size scaling \eqref{FiniteSize} and the fact that $\nu = 1/(2-\zeta)$~\cite{Narayan1993}, we rescale $(f_c - f_{\infty})$ by $\gamma$. As seen in Fig.~\ref{qEW}(b), the data collapse is very good.
Furthermore, that the correct finite size scaling laws are recovered for sufficiently large targets is further evidenced by the measured scaling of the standard deviation of $f_c(L)$, as plotted in Fig.~\ref{qEWStdDev}. This has the asymptotic form $L^{-1/\nu}$, as expected from finite size scaling.

\begin{figure}
\begin{center}
\includegraphics[width=0.9\columnwidth]{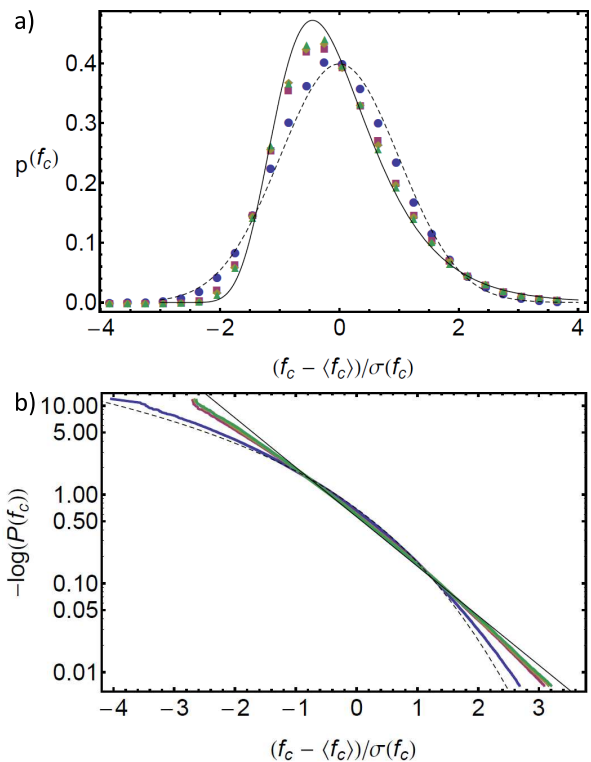}
\end{center}
\caption{(a) Histograms of the measured depinning force over $10^5$ trials, for $L=100$, with $\gamma=0.01$ (circles), $1$  (squares), $10$ (diamonds) and $100$ (triangles). The distributions have been normalized to have zero mean and unit variance. Error bars are smaller than the symbol size. (b) Cumulative distribution functions, showing convergence towards the Gumbel distribution (a straight line on this double-log plot) as $\gamma$ increases.  For comparison, in both plots, the Gaussian distribution (dashed line) and Gumbel distribution (solid thin line) are also shown. }
\label{fcDistribution}
\end{figure}

We can go beyond the moments of the distribution of $f_c$ to consider the distribution itself. It has previously been shown~\cite{Bolech:2004, Fedorenko2006} that the distribution of $f_c$ measured in a finite sample of size $L\times M$ (with open boundary conditions in the $M$ direction) depends on the ratio of $M$ to the interface width $w L^{\zeta}$. Heuristically, the system can be divided into $n=M/(w L^{\zeta})$ subsystems which are approximately independent, each with its own depinning force. The depinning force in the entire $L\times M$ system is the maximum of these, and if $n\to\infty$ the measured depinning force is described by extreme value statistics, in particular, the Gumbel distribution, which has cumulative distribution $P(x)=\exp(-e^{-(x-\mu)/\beta)})$, where $\mu$ is the modal value of $x$ and the standard deviation of the distribution is $\beta \pi/\sqrt{6}$. In Fig.~\ref{fcDistribution}, we plot the distribution of $f_c$  (rescaled to have zero mean and unit standard deviation) for $L=100$, for a range of $\gamma$ values. The targets range from a few interface widths to $\sim10^4$ interface widths. In agreement with Refs.~\cite{Bolech:2004, Fedorenko2006}, as $\gamma$ increases, the distribution approaches the Gumbel limit, albeit very slowly in the tails of the distributions.

\section{Other simple depinning models: KPZ and mean field}
\label{MFSect}
\begin{figure}
\begin{center}
\includegraphics[width=0.9\columnwidth]{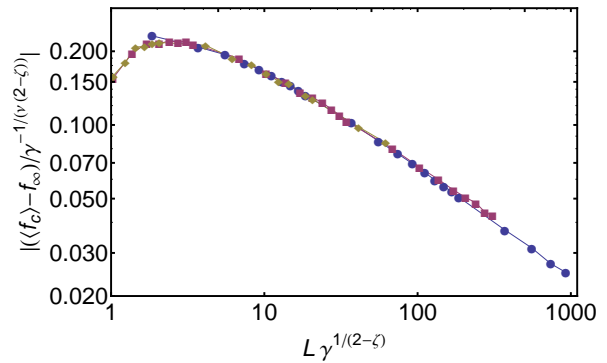}
\end{center}
\caption{Data collapse for the KPZ model, based on the same arguments as the quenched Edwards-Wilkinson model (Eq.~\ref{FiniteSize} and \ref{TargetGreaterWidth}), using fitted parameters $\zeta=0.635$, $f_{\infty}=0.226$, $1/\nu=0.55$. The symbols represent $\gamma=0.1$ (circles), $\gamma=0.01$ (squares) and $\gamma=0.005$ (diamonds). The standard error is smaller than the symbol size.}
\label{figKPZ}
\end{figure}

The role of the parameter $\gamma$ relative to the interface width does not depend specifically on the choice of interaction $f_{\rm el}$ used in the quenched Edwards-Wilkinson equation, and we have verified that models from other universality classes are affected by the choice of $\gamma$ in the same way. 
For example, in the KPZ model~\cite{Kardar1986}, where the elastic force $f_{\rm el}=\nabla^2 u(x)$ is supplemented by a nonlinear term proportional to $(\nabla u)^2$, we measure a roughness exponent of $\zeta=0.635$ (consistent with other measurements~\cite{Rosso2003, Chen2011}) and a finite size exponent of $1/\nu \approx 0.55$.  Per Eq.~\eqref{TargetGreaterWidth}, $f_c(L)$ curves for different $\gamma$ can be collapsed via the rescaling $L\to \gamma^{1/(2-\zeta)}L$, $(f_c-f_{\infty}) \to  \gamma^{-1/(\nu(2-\zeta))} (f_c-f_{\infty})$, as shown in Fig.~\ref{figKPZ}. 

The quenched Edwards-Wilkinson and KPZ models both have local interactions, and the interface width scales with length as $L^{\zeta}$. We now turn to a mean-field model~\cite{Fisher1983, Fisher1985, Vannimenus2001}, in which every site interacts with every other site, and the interface width does not depend on the length scale it is measured over ($\zeta=0$). 
We couple the interface to its center of mass, $\bar{u}$. The total force on site $x$ is
\begin{equation}
f_x = f + J(\bar{u} - u_x)+ \eta_{u, x}.
\end{equation}

\begin{figure}
\begin{center}
\includegraphics[width=\columnwidth]{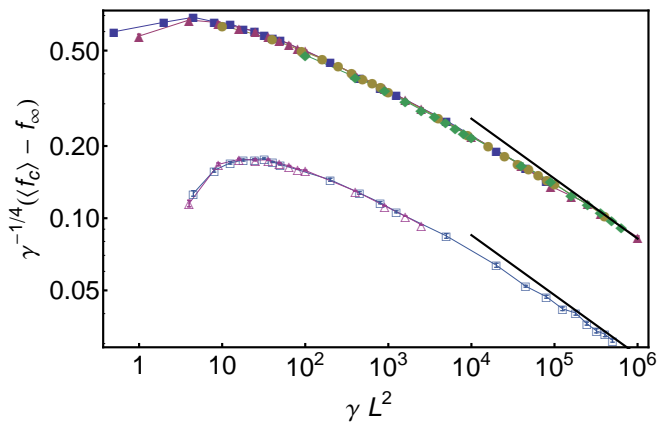}
\end{center}
\caption{Rescaled depinning force \textit{vs} target displacement $\gamma L^2$, in the mean field model. The solid black lines are proportional to $L^{-1/2}$. Closed symbols represent data for $J=1$, open symbols represent data for $J=0.1$, the $\gamma$ values are $0.005$ (squares), $0.01$ (triangles), $0.1$ (circles) and $1$ (diamonds). For $J=0.1$, $f_{\infty}\approx0.64$, for $J=1$, $f_{\infty}\leq10^{-3}$.  }
\label{MFUnscaledJ}
\end{figure}

In Fig.~\ref{MFUnscaledJ} we show the finite size behavior of the depinning force  for this model. The finite size exponent takes the mean-field value of $\nu=2$. For fixed $J$ we can collapse data from simulations with different $\gamma$ values following the same argument as for the quenched Edwards-Wilkinson model, but with the interface width constant. As seen in Fig.~\ref{MFUnscaledJ}, the data collapse is excellent.

The coupling strength $J$ affects the finite size scaling in two ways. First, as $J$ increases, the interface becomes flatter. Accordingly, as seen in Fig.~\ref{MFUnscaledJ}, for larger coupling strengths, the peak in $f_c(L)$ occurs for smaller targets. The second effect of $J$ is to alter the depinning threshold, so that as $J$ increases, $f_{\infty}$ decreases. Heuristically, this can be understood by considering that as $J$ increases, so does the forward force on sites lagging behind the mean interface position, so that the external driving force needs only be able to overcome the local pinning force on a few sites to start a large avalanche. 

For completeness, we have also made simulations using Gaussian distributed quenched disorder, with $\langle \eta_i \rangle = 0.5, \sigma(\eta_i) = 0.2$, and find very similar behavior to that observed with uniform-distributed disorder.

\section{Effects of hardening and machine stiffness}
\label{hardening}

Finally, we discuss two related scenarios: strain hardening and strain-controlled experiments.
In materials subject to hardening, the external stress required to induce plastic deformation grows with the plastic strain. A simple model for this can be incorporated into the model for interface depinning, by adding a force term that is opposed to the interface motion, and proportional to the strain associated with the interface displacement, $\bar{u}/L^2$, where $\bar{u}$ is the average of $u_x$ over $x$. The total force on site $x$ is then
\begin{equation}
\label{eqHardening}
f_x =  f_{\rm el}(x, u_x) - \eta(x, u_x) + f - k \frac{\bar{u}}{L^2},
\end{equation}
where $k$ is a hardening constant. The hardening force, $f_H=-k\bar{u}/L^2$, effectively reduces the external force acting on the dislocation as the dislocation displacement increases, and for large displacements the hardening dominates and the driving force required to move the interface is proportional to the mean displacement. As a result, the interface remains critical and never depins.  

Alternatively, in the case of a strain controlled experiment, one can imagine that a subsystem undergoing plastic deformation is coupled to a spring of stiffness $k$, and a total deformation $\gamma_T$ is imposed (as illustrated in the Inset of Fig.~\ref{StressStrain}(b)). In the absence of a plastic subsystem, $\gamma_T$ would be attained using a stress of $\sigma_T=k\gamma_T$, per Hooke's law, but when plastic deformation occurs, the spring extension/compression is reduced by $\bar{u}/L^2$ so that $\sigma_T$ is given by
\begin{equation}
\label{StrainControlled}
\sigma_T = k (\gamma_T - \frac{\bar{u}}{L^2}).
\end{equation}
Simulations of an interface subject to strain hardening can be reinterpreted in terms of strain controlled experiments~\cite{Zaiser2005, Zaiser2007}, by treating $f$ as $k \gamma_T$ and plotting $\sigma_T$ \textit{vs} $\gamma_T$, as illustrated in Fig.~\ref{StressStrain}(b). We work in units where stresses and forces are interchangeable.

We will first discuss how size effects in interface depinning are affected by strain hardening, and then discuss size effects in strain controlled experiments.

\subsection{Hardening}
We have simulated the quenched Edwards-Wilkinson model with an added hardening term. A typical ``stress-strain'' curve is shown in Fig.~\ref{StressStrain}(a). For small driving force and interface displacement, the behavior is that of the quenched Edwards-Wilkinson model without hardening, but as the driving force and displacement are increased, the ``stress-strain'' curve departs from its zero-hardening counterpart. When hardening dominates, if a force of $f^*$ has been applied to drive the interface to position $\bar{u}^*$ and it has come to a stop, the force required to start it moving again is
\begin{equation}
\label{StressStrainHardening}
f = f^* + k \frac{\bar{u}^*}{L^2}.
\end{equation}
This is the behavior seen in the ``stress-strain'' curve, which approaches a line of slope $k/L^2$ in the limit $\bar{u}\to\infty$.

\begin{figure}
\begin{center}
\includegraphics[width=0.9\columnwidth]{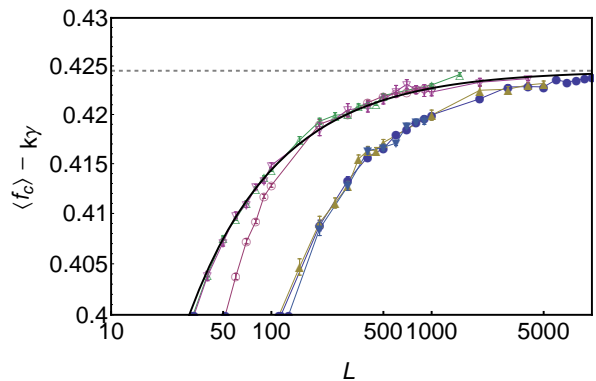}
\end{center}
\caption{Size dependence of depinning force for the quenched Edwards-Wilkinson equation subject to hardening of strength $k$. For sufficiently large $k \gamma L^2$, data from simulations with different $\gamma$ values can be collapsed by subtracting $k \gamma$ from $\langle f_c \rangle$, to give finite size scaling of the form described by Eq.~\eqref{FiniteSizeHardening}, as indicated by the black solid line. This collapse does not work for small $k\gamma L^2$, as seen in the data for $k=1, \gamma=0.01$ (open circles), which only fall on to the collapsed curve for sufficiently large $L$. The dashed horizontal line indicates $f_{\infty}$, the $L\to\infty$ limit of $f_c(L)$ in the absence of hardening. Solid symbols represent $k=10$ and open symbols represent $k=1$; the shapes represent $\gamma=0.01$ (circles), $\gamma=0.15$ (triangles) and $\gamma=2$ (inverted triangles). Error bars represent the standard error.}
\label{hardeningFig}
\end{figure}

In Fig.~\ref{hardeningFig}, we plot typical $f_c$ \textit{vs} $L$ curves. With the introduction of hardening, it is no longer possible to collapse data for the whole space of parameters $\gamma$ and $k$. Instead, the observed behavior depends on $k \gamma$.
For large $k \gamma$, $f_c(L)$ curves from simulations with the same $k$ can be collapsed by subtracting $k \gamma$ from $\langle f_c \rangle$, but data from simulations with different $k$ cannot be collapsed onto a single form. Furthermore, if the $k \gamma$ is small, this procedure does not yield data collapse, as exemplified in Fig.~\ref{hardeningFig} by the $f_c(L)-k\gamma$ curve for $\gamma=0.01, k=1$.

We can interpret this by considering the maximum hardening force applied during simulations, which is applied when the interface is at its target: $\bar{u}=\gamma L^2$. For small $k \gamma$, the hardening force is always negligible, and the size effects are the same as those discussed in previous Sections. 
When $k \gamma$ is large, hardening becomes important. A first approximation to understand its role is to imagine that the interface is driven to the target without hardening, and then the hardening force is turned on. Assuming the target is sufficiently large, the force required to attain it (with zero hardening) is given by Eq.~\eqref{FiniteSize}. Then, from Eq.~\eqref{StressStrainHardening}, the force required to continue to move the interface is
\begin{equation}
\label{FiniteSizeHardening}
f =f_{\infty} + a L^{-1/\nu} + k \gamma,
\end{equation}
where $f_{\infty}$ is the $L\to\infty$ limit of $f_c(L)$ when $k=0$. This form agrees well with our simulation data, as seen in Fig.~\ref{hardeningFig}.
Because of the extra $\gamma$-dependent term, it is not possible to use linear rescalings to eliminate the $\gamma$ dependence of $(f_c - f_{\infty})$, and exact data collapse is not possible.

\subsection{Strain controlled experiments}
\begin{figure}
\begin{center}
\includegraphics[width=0.9\columnwidth]{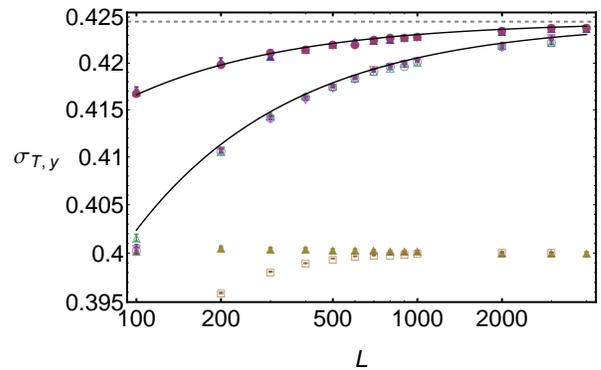}
\end{center}
\caption{Size dependence of the measured yield stress $\sigma_{T,y}$ in strain controlled simulations of the quenched Edwards-Wilkinson model, with spring constant $k=10$ (open symbols), and $k=1$ (filled symbols). The yield strains used are $\gamma_{T, y}=1$ (triangles), $0.5$ (circles), $0.4$ (inverted triangles) and $0.04$ (squares). The black lines indicate the expected finite size scaling of Eq.~\eqref{FiniteSizeStrainControlled}, with $f_{\infty}=0.424$ and $1/\nu=0.75$. When $k \gamma_{T,y}$ is too small, this scaling breaks down because the interface never reaches criticality, as seen for the lower two curves ($k=10, \gamma_{T,y}=0.04$, and $k=1, \gamma_{T,y}=0.4$).}
\label{strainControlled}
\end{figure}

In strain controlled simulations, the choice of yield strain is more crucial than in stress controlled simulations. In Fig.~\ref{strainControlled}, we plot the size dependence of $\sigma_{T,y}$ for several choices of yield strain $\gamma_{T, y}$, for strain controlled simulations of an interface described by the quenched Edwards-Wilkinson equation. For sufficiently large $\gamma_{T, y}$, the behavior described in Eq.~\eqref{FiniteSize} is recovered and $\sigma_{T,y}(L)$ approaches the same $L\to\infty$ limit found in stress controlled simulations, with the same finite size exponent $1/\nu=0.75$. However, for smaller $\gamma_{T, y}$, this picture breaks down.

We can interpret this as follows. Recall that in stress controlled simulations of the quenched Edwards Wilkinson model, the target $\gamma L^2$ grows faster than the interface width $w L^{\zeta}$ and therefore as $L\to\infty$ the correct finite size scaling must occur.
However, in strain controlled simulations, the target $\gamma_{T, y}=f_y/k$ may correspond to a driving force $f_y$ that is always smaller than the true depinning force (as given in Eq.~\eqref{FiniteSize}). In that case, the interface never reaches criticality for any $L$, and the standard finite size scaling does not hold. Note that the minimum strain target depends on both the depinning force of the interface and also the spring constant $k$, as shown in Fig.~\ref{strainControlled}.

If the target is large enough for the interface to reach criticality, then the size effects can be predicted by re-interpreting the strain controlled simulation as a stress controlled simulation with hardening. The finite size behavior of $f=k\gamma_T$ is given by  Eq.~\eqref{FiniteSize} and \eqref{StressStrainHardening}, so that
\begin{equation}
\label{FiniteSizeStrainControlled}
\sigma_{T,y} = f-k\frac{\bar{u}_y}{L^2} = f_{\infty} + a L^{-1/\nu},
\end{equation}
where $\bar{u}_y$ is the mean interface displacement when the total strain is equal to the yield strain. In other words, for sufficiently large yield strains, stress controlled and strain controlled simulations give the same results, and furthermore, in that case, the measured yield stress in a strain controlled experiment does not depend on the yield strain.

\section{Conclusion}
The question of how to measure yield stress is an important methodological problem in the determination of finite size effects in microscale plasticity, which to our knowledge has received little attention previously. In this paper, we have gained insight into how to avoid spurious size effects in a simplified model for the onset of plastic yield when the yield strain is arbitrarily defined. Reassuringly, in all the models we study, for sufficiently large systems the correct finite size scaling is recovered. However, for smaller systems, the finite size effects are not described by the scaling law~\eqref{FiniteSize}, and instead depend on various details of the model under study, a result which has implications especially for simulation studies of depinning and plasticity, where until recently only relatively small sizes were accessible computationally.

We also observe that the idea that the crossover between ``small'' and ``sufficiently large'' systems is determined by the roughness of a self-affine interface is one that has broader relevance. For example, experimental studies of polycrystalline copper samples under tension~\cite{Zaiser2004} have revealed that profiles of the surface height $y(x)$(with $x$ along the sample axis) are self-affine, that is, $W(L)\equiv \langle|y(x+L)-y(x)|\rangle\sim L^H$, with $H \approx 0.75$ for imposed strains $\gtrsim 10\%$. This surface profile is related to the variation in the strain, which in a sample of length $L$ is $\Delta \gamma = W(L)/L \sim L^{-0.25}$. Imposing the requirement that the total strain should be larger than its variance, a minimum system size $L_c$ is determined by setting $\gamma = \Delta\gamma$. In the case of polycrystalline copper, this length scale turns out to be small enough to have no experimental effect: using the data from Zaiser \textit{et al.}'s experiments~\cite{Zaiser2004}, $L_c \approx 1.6\times10^{-9}$~m for $\gamma_{\rm tot}=10\%$. For different materials and loading conditions, however, it may be possible for $L_c$ to be comparable with experimental sample sizes, leading to intricate size effects.

\begin{acknowledgments}
This work is supported by the European Research Council through the Advanced Grant 2011 SIZEFFECT. We thank Michael Zaiser, Zsolt Bertalan and Gianfranco Durin for helpful discussions.
\end{acknowledgments}

\end{document}